\documentclass{aa}  

\usepackage{graphicx}	
\usepackage{amsmath}	
\usepackage{xcolor}
\usepackage{hyperref}
\graphicspath{{figures/}}

\usepackage{xspace}

\newcommand{\hi}{HI}

\usepackage{txfonts}
\def\be{\begin{equation}} 
\def\ee{\end{equation}} 
\def\ba{\begin{eqnarray}} 
\def\ea{\end{eqnarray}}

\def\msun{{\Msun}}

\newcommand{\NHI}{\ifmmode{N_\mathrm{HI}}\else $N_\mathrm{HI}$\xspace\fi}

\def\gsim{\lower.5ex\hbox{\gtsima}} 
\def\lsim{\lower.5ex\hbox{\ltsima}} \def\gtsima{$\; \buildrel > \over 
\sim \;$} \def\ltsima{$\; \buildrel < \over \sim \;$} \def\prosima{$\; 
\buildrel \propto \over \sim \;$} \def\gsim{\lower.5ex\hbox{\gtsima}} 
\def\lsim{\lower.5ex\hbox{\ltsima}} 
\def\simgt{\lower.5ex\hbox{\gtsima}} 
\def\simlt{\lower.5ex\hbox{\ltsima}} 
\def\simpr{\lower.5ex\hbox{\prosima}}   
  
 \def\gtsima{$\; \buildrel > \over \sim \;$} 
\def\ltsima{$\; \buildrel < \over \sim \;$} 
\def\gsim{\lower.5ex\hbox{\gtsima}} 
\def\lsim{\lower.5ex\hbox{\ltsima}} 
\def\simgt{\lower.5ex\hbox{\gtsima}} 
\def\simlt{\lower.5ex\hbox{\ltsima}} 
\def\simpr{\lower.5ex\hbox{\prosima}}

\newcommand{\Lya}{\ifmmode{\mathrm{Ly}\alpha}\else Ly$\alpha$\xspace\fi}

\def\msun{\,{\rm \Msun}}

\def\E3{{\cal E}_{\rm g}^{III}}

\def\r12{r_{1/2}} 
\def\x12{x_{1/2}} 
\def\v12{v_{1/2}}

\newcommand\code[1]{\textsc{\MakeLowercase{#1}}}

\def\nh2{n_{\rm H2}}
\def\fh2{f_{\rm H2}}

\def\arcsec{^{\prime\prime}}
\def\angstrom{\textrm{A\kern -1.3ex\raisebox{0.6ex}{$^\circ$}}}

\def\msun{{\rm M}_{\odot}}

\def\msunyr{\msun\,{\rm yr}^{-1}}

\def\highz{$\mbox{high-}z$~}

\begin{document}

   \title{Neutral hydrogen in and around galaxies during the Epoch of Reionization}

   \author{V.~Gelli
        \inst{1,2}
        \and
        C.~Mason\inst{1,2}
        \and
        A.~Pallottini\inst{3,4}
        \and
        K.~E.~Heintz\inst{1,2,5}
        \and
        Z.~Chen\inst{1,2}
        \and
        V.~D'Odorico\inst{6,7}
        \and
        A.~Ferrara\inst{4}
        \and
        J.~Fynbo\inst{1,2}
        \and
        M.~Kohandel\inst{4}
        \and
        C.~L.~Pollock\inst{1,2}
        \and
        C.~Robinson\inst{1,2}
        \and
        S.~Salvadori\inst{8,9}
        }

   \institute{Cosmic Dawn Center (DAWN)\\
              \email{viola.gelli@nbi.ku.dk}
        \and
             Niels Bohr Institute, University of Copenhagen, Jagtvej 128, 2200 København N, Denmark
        \and
             Dipartimento di Fisica ``Enrico Fermi'', Universit\'{a} di Pisa, Largo Bruno Pontecorvo 3, Pisa I-56127, Italy
        \and
             Scuola Normale Superiore, Piazza dei Cavalieri 7, I-56126 Pisa, Italy
        \and
            Department of Astronomy, University of Geneva, Chemin Pegasi 51, 1290 Versoix, Switzerland
        \and
            INAF - Osservatorio Astronomico di Trieste, Via Tiepolo 11, 34143 Trieste, Italy
        \and
            IFPU - Institute for the Fundamental Physics of the Universe, via Beirut 2, I-34151 Trieste, Italy
        \and
            Dipartimento di Fisica e Astrofisica, Università degli Studi di Firenze, Via G. Sansone 1, I-50019, Sesto Fiorentino, Italy
        \and
            INAF - Osservatorio Astrofisico di Arcetri, Largo E. Fermi 5, 50125, Firenze, Italy
             }

   \date{Received XXX; accepted XXX}

 
  \abstract
   {JWST spectra revealing Lyman-$\alpha$ (\Lya) absorption in $z \sim 5-14$ galaxies offer a unique opportunity to probe the earliest stages of reionization. However, disentangling absorption by the increasingly neutral intergalactic medium (IGM) from that in galaxies’ interstellar and circumgalactic medium (ISM, CGM) remains challenging due to the poorly constrained nature of gas in and around galaxies at these redshifts.}
   {We use cosmological zoom-in simulations to characterize the distribution and redshift evolution of neutral hydrogen (\hi) along sightlines to star-forming regions during the Epoch of Reionization, to interpret the contribution of local \hi\ (ISM+CGM) to \Lya absorption in $z>5$ galaxy spectra.
   }
   {We analyze $\sim$100 $z = 6 - 9.5$ galaxies from the high resolution \code{serra} simulations, which have been shown to reproduce $z>6$ cold ISM properties, generating mock sightlines from star formation peak for each galaxy. We explore the resulting sightline distribution of \hi\ column densities (\NHI) and its variation with radius, halo mass, and redshift.}
   {We find broad sightline variation in \NHI\ (0.5–1.5~dex) arising from complex ISM morphology driven by bursty star formation, with median, $\log(\NHI/{\rm cm}^{-2})\simeq21$–22. 
   Dense gas in galaxies' ISM is the dominant origin of damped \Lya absorption (DLA) systems along sightlines towards star-forming regions, outweighing gas in the CGM, filaments, and proximate absorbers. 
   Median \NHI\ increases with halo mass, scaling approximately with the virial radius, as expected due to larger potential wells and more extended CGM, but shows no significant redshift evolution at fixed halo mass. 
   We suggest this implies \NHI distributions measured at the end of reionization may provide useful priors for interpreting IGM damping wings at higher redshifts.
   We investigate $\NHI > 10^{22}$\,cm$^{-2}$ sightlines to interpret strong $z > 5$ galaxy-DLA candidates identified with JWST. We find these trace dense, metal-enriched, neutral ISM within $<1$\,kpc of massive halos ($M_h \gtrsim 10^{11}\,M_\odot$), a scenario that can be tested with higher-resolution spectroscopy.
   }
   {}

\maketitle

\section{Introduction}

The scattering of Lyman-$\alpha$ (\Lya) photons is a powerful probe of neutral hydrogen in the universe \citep[see e.g.,][for reviews]{Dijkstra2014, Ouchi2020}. The observed decrease in transmission of \Lya photons emitted by quasars and galaxies at $z\gtrsim5$ provides some of our strongest evidence for the transition of intergalactic hydrogen from neutral post-Recombination to almost fully ionized today \citep[e.g.,][]{Fan2006,Davies2018b,Mason2018,Bosman2022,Tang2024c}.
When the mean neutral hydrogen fraction, $\overline{x}_{\rm HI}$, in the intergalactic medium (IGM) exceeds $\simgt 20-30\%$, large, diffuse neutral hydrogen patches, spanning tens of comoving Mpc, are expected to still persist. These are large enough to produce high optical depth even in the Lorentzian ``damping'' wings of the \Lya absorption profile: scattering photons emitted up to thousands of km/s redward of \Lya line center.  
Thus a key observational signature of a highly neutral IGM is smooth suppression of flux redward of \Lya in the continuum of high redshift sources \citep{Miralda-Escude98, mcquinn_studying_2007}. 

Prior to JWST, these \Lya IGM damping wings had been observed in a handful of bright quasars at $z\sim6-7.5$ \citep{Mortlock2011,Banados2018,Durovcikova2019,Wang2020,Yang2020,Greig2024}, and in an integrated sense in the declining \Lya equivalent width distribution and \Lya luminosity functions in galaxies at $z\sim6-8$ \citep[e.g.,][]{Stark2011,Treu2013,Pentericci2014,Ouchi2017,Mason2019c,jung_texas_2020,Bolan22}. Comparison of these observations with damping wing sightlines from reionization simulations imply the IGM is $\sim50\%$ neutral at $z\sim7-8$, consistent with CMB constraints \citep[e.g.,][]{Greig19,Davies2018b,Mason2018,Morales2021}.

Excitingly, JWST has opened a new window on galaxies, and the impact of the IGM on their spectra, deep into the earliest stages of reionization at $z>10$ \citep[see e.g.][for recent reviews]{Adamo24_review, Stark2025}. NIRSpec's unprecedented sensitivity \citep{Jakobsen202, Rigby2022} has enabled detections of \Lya emission in UV faint galaxies and the first deep spectroscopy of the UV continuum of star-forming galaxies at $z\simgt5-14$. 
Early JWST spectroscopy of $z\simgt8$ galaxies revealed apparent \Lya damping wings in several \highz galaxies \citep[e.g.][]{CurtisLake22,Hsiao2024,Heintz2023b}. Larger censuses of hundreds of $z\sim5-14$ galaxies have shown a systematic decrease in flux around the \Lya break with increasing redshift \citep[][]{Umeda2024,Umeda2025,Asada2024,Heintz25,Mason25}, which implies a predominantly neutral IGM by $z>9$ \citep{Umeda2024,Mason25}. Simultaneously, JWST measurements of \Lya emission lines have revealed a continued decline in strong \Lya emission at $z>8$ \citep[e.g.,][]{Nakane24,Jones2024_Lya,Tang2024c}. These observations showcased the potential of JWST for advancing our understanding of the early stages of reionization.

However, to fully exploit the capabilities of JWST to infer the imprint of the IGM on distant galaxies requires an understanding of galaxies' spectra \textit{prior} to attenuation by the neutral IGM. 
In particular, dense neutral hydrogen in the interstellar and/or circumgalactic (ISM, CGM) medium of galaxies can also produce \Lya damping wing absorption. 
Although the damping wings produced by this local \hi\ and the IGM differ in their spectral shape \citep{Mesinger2008,Lidz2021}, disentangling them can be challenging with JWST’s low-resolution prism spectra \citep[][]{Umeda2024,Mason25,Huberty2025}.
In particular, local absorption systems in the ISM and CGM with column densities $\NHI > 10^{20.3}$\,cm$^{-2}$ (i.e. damped \Lya absorbers, DLAs) produce opacity at $\gtrsim 1220  \mathring{\rm A}$ that would be visible in the continuum of galaxies observed by JWST.

\begin{figure*}[t!]
\centering
\includegraphics[width=\textwidth]{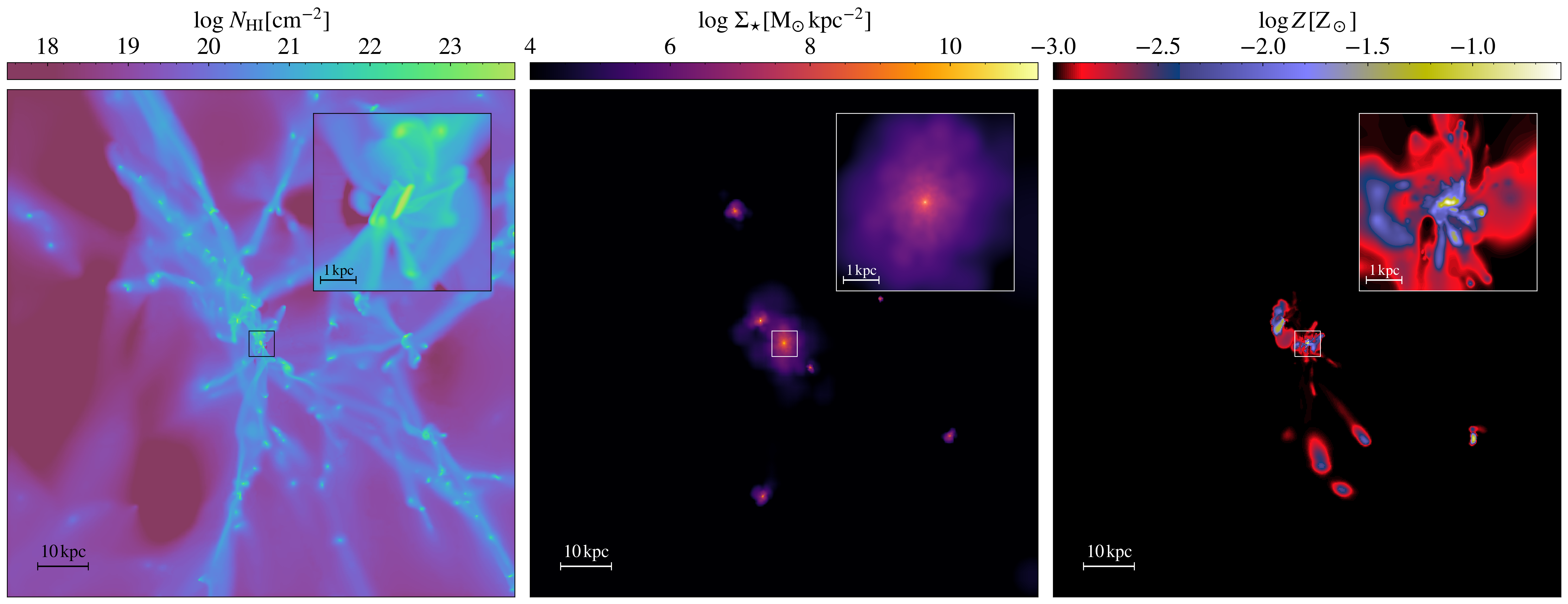}
\caption{Maps of \textit{Amana}, typical star forming $z=7$ galaxy of $M_\star \sim 10^{9.5}\msun$ and ${\rm SFR}\sim 15~\msunyr$ from the \code{serra} simulation. The three panels show quantities along the $z$ direction: the neutral hydrogen column density ($N_{HI}$), the stellar mass surface density ($\Sigma_\star$), and the sightline mass-averaged gas metallicity ($Z$). Each map covers a 100~kpc field of view, with an inset showing the ISM by zooming on the inner 5~kpc.
\label{fig:z8_galaxy}
}
\end{figure*}

At lower redshifts, multiple observations have shown that galaxies contain dense \hi\ gas. Spectroscopy of star-forming galaxies at $z\sim 0-5$ indicates absorption around \Lya is common, often in addition to \Lya emission \citep[e.g.,][]{Shapley2003, Reddy2016, McKinney19, Pahl2023,Hu23}; with \Lya emission typically offset by $\sim200-300$\,km/s from the systemic redshift \citep[e.g.,][]{Erb2014,Steidel2014,Shibuya2017,Prieto-Lyon2023b}; furthermore, integral field observations have revealed \Lya emission halos, extending beyond galaxies' UV emission \citep[e.g.,][]{Fynbo03, Steidel2011,Momose2014,Wisotzki2016,Leclercq2017}. These observations all imply scattering of \Lya photons by high column density neutral gas, $\NHI \sim 10^{20-22} \rm cm^{-2}$, which would be classified as DLAs, in galaxies' ISM and CGM.
DLAs have also been widely studied in quasar spectra, and have been associated with foreground galaxies and their surrounding environments \citep[$<100$~ proper (p)kpc, e.g.][]{Rudie13, Turner14}, with the highest \hi\ column densities correlating with the smallest impact parameters to galaxies \citep[e.g.,][]{Rubin2015, Noterdaeme14, Krogager20}.
DLAs have also been detected in long-duration gamma ray bursts (GRB) spectra at $z\sim2-6$ \citep[e.g., ][]{Jakobsson2006, Prochaska07, Fynbo09, Tanvir19, Heintz23_grb}. GRB-DLAs typically have higher column densities ($\NHI > 10^{21} \rm cm^{-2}$) and metallicities than those observed in the spectra of background quasars \citep{Jakobsson2006,Prochaska07,Fynbo2008}.
This is consistent with GRB-DLAs arising from dense \hi\ in the ISM, within $<1$~kpc from the center of the host galaxy, as GRBs are associated with the death of massive stars \citep[e.g.,][]{Blooom2002,Pontzen10, Krogager2024}.

JWST has now extended \Lya absorption studies in individual galaxies out to $z\sim14$, finding a very wide range of \Lya break shapes \citep[e.g.,][]{Umeda2024,Heintz25,Mason25}. 
While the inferred median local HI column densities are comparable to those seen in $z\sim3$ galaxy stacks \citep[$\NHI\sim10^{21}$\,cm$^{-2}$][]{Reddy2016} there are some sources where the observed absorption significantly exceeds that predicted by the IGM alone \citep[estimated to be $\sim20$\% of sources at $z\sim5-13$][]{Mason25}, implying DLAs with $\NHI \simgt 10^{22} \rm cm^{-2}$ \citep[e.g.,][]{Hsiao2024,Heintz2023b,Chen24,Umeda2024,Carniani2025,Mason25}.
While the origin of these high inferred column densities has been debated\footnote{In some sources, which also show Balmer jumps and strong nebular emission lines, \citet{Cameron24} and \citet{Katz2024} have argued two-photon nebular continuum emission may provide a better explanation for strong turnovers seen around the \Lya break.}
\citep[e.g., large gas reservoirs in the CGM, dense infalling filaments, foreground protoclusters, dense ISM,][]{Heintz2024b,Terp2024,Chen24}, they are comparable to the highest column densities measured in GRB sightlines \citep[e.g.,][]{Jakobsson2006,Watson2006,Prochaska09,Tanvir19,Saccardi2025} which suggests they also arise from dense gas in the ISM.
At these extreme column densities, local absorption can dominate over the IGM damping wings. Thus, to better interpret JWST spectra in the context of reionization, it is important to identify the origin of high column density \hi\ gas in the ISM + CGM of galaxies and to understand how its prevalence may evolve with galaxy properties and redshift.

Cosmological hydrodynamic simulations provide a powerful tool to investigate the expected properties of \highz environments. 
There has been significant focus on using simulations to interpret DLAs observed in quasar spectra, which probe gas at varying transverse distances from foreground galaxies \citep[e.g.,][]{Haehnelt1998,Pontzen2008,Rahmati15,pallottini:2014b}. These studies have investigated the covering fractions of \hi\ gas in the CGM of galaxies, finding, in agreement with observations, higher \NHI\ values at lower impact parameters, i.e., when quasar sightlines intercept more central regions of foreground galaxies \cite[e.g.,][]{Faucher-Giguere15, Rahmati15, Tortora24}.
However, these studies have primarily focused on analyzing sightlines passing through the galaxy ISM and CGM but originating from background QSOs.  
Interpreting JWST observations requires a new approach: we need to investigate sightlines directly \textit{towards} the central star-forming regions in the galaxies themselves, capturing the impact of local \hi\ structures along the line-of-sight to the H\,{\sc ii} regions which dominate most JWST galaxy spectra. 
\citet{Pontzen10} performed a comparable analysis in the context of GRB-DLAs, to investigate \NHI in sightlines to star-forming regions using simulations at $z=3$. 
To interpret JWST observations, it is now necessary to extend such an analysis to $z>5$ and explore how the distribution of high column density gas around star-forming regions depends on galaxy properties and redshift during the Epoch of Reionization.

In this paper, we use the \code{serra} zoom-in simulations \citep{Pallottini2022} to investigate the nature and redshift evolution of \highz \hi\ absorbers along sightlines to star-forming regions, focusing on understanding the origin and evolution of gas responsible for the damped \Lya absorption in galaxy spectra (i.e. $\NHI \gtrsim 10^{20}$\,cm$^{2}$).
Where is the highest \hi\ column density gas located? Does this gas primarily trace dense star-forming regions in the ISM, dense inflowing filaments or gas rich low-mass halos in the more extended CGM, or chance intersections through other galaxies? What column densities are expected, and how do they evolve with cosmic time and galaxy properties?

The paper\footnote{Throughout the paper we assume a $\Lambda$CDM model with vacuum, matter, and baryon densities in units of the critical density $\Omega_{\Lambda}= 0.692$, $\Omega_{m}= 0.308$, $\Omega_{b}= 0.0481$, Hubble constant $\rm H_0=67.8\, {\rm km}\,{\rm s}^{-1}\,{\rm Mpc}^{-1}$, spectral index $n=0.967$, and $\sigma_{8}=0.826$ \citep{Planck14}. All spatial coordinates and distances are expressed in physical units.} is organized as follows: in Sec.\ref{sec:simulation}, we describe the simulations and methods used; in Sec.\ref{sec:results}, we present the properties of neutral gas around typical \highz galaxies and how these depend on galaxy properties and their evolution; in Sec.\ref{sec:discussion}, we discuss the implications and limitations of our results in the context of previous simulations and provide observational predictions; and in Sec.\ref{sec:conclusions}, we summarize our main conclusions.

\section{The simulations} \label{sec:simulation}

We adopt the \code{SERRA} suite of high-resolution cosmological zoom-in simulations \citep{Pallottini2022}. It is designed to model the evolution of typical Lyman-break galaxies during the Epoch of Reionization, successfully reproducing observed ISM properties \citep[e.g.,][]{Kohandel24, Kohandel25,  Gelli25, Vallini2025} thanks to its treatment of non-equilibrium chemistry and high resolution ($\simeq 30 \rm pc$) that resolves star-forming regions down to molecular cloud scales.

The simulations are initialized at $z=100$ using the \code{MUSIC} code \citep{HahnAbel11} to generate initial conditions within a cosmological volume of $(20 \, {\rm Mpc}/h)^3$, adopting the cosmological parameters from \citep{Planck14}.
After a target dark matter (DM) halo is selected, the simulations zoom-in by following the evolution of DM, gas, and stars, which are simulated using a customized version of the \code{RAMSES} Adaptive Mesh Refinement code \citep{Teyssier02}, achieving a baryonic mass resolution of $1.2 \times 10^4 \msun$ and a spatial resolution reaching $\sim30$~pc in the zoom-in regions.

Radiative transfer is incorporated on-the-fly via \code{RAMSES-RT} \citep{Rosdahl2013}, where photons are divided into 5 energy bins in the range $6.0-24.59\rm~eV$ tracked separately. The cosmic UV background (UVB) contribution is neglected as the typical ISM densities are large enough to ensure self-shielding. In the EoR and down to $z\sim 5$, statistically, the ionization from the UVB is negligible for gas with density $n\gtrsim10^{-2}$ \citep{Rahmati13, Chardin2015} as local sources dominate the photon budget.
The interaction between gas, dust, and photons is regulated using a non-equilibrium chemical network generated with \code{KROME} \citep{grassi:2014,Pallottini16}. This includes the individual tracking of $\rm H, H^+, H^-, He, He^+, He^{++}, H_2,H_2^+$ and electrons, and a total of 48 reactions \citep{Bovino16, Pallottini17}.
The formation of molecular hydrogen, fundamental in regulating star formation, happens in both the gas-phase and on dust grain surfaces \citep{Jura74}.
For the dust, we assume that the dust-to-gas mass ratio scales with metallicity. To account for the unresolved effects of the first stars and mini-halos, the initial gas metallicity is set to a floor value of $Z_{\text{floor}} = 10^{-3} \rm Z_\odot$ \citep{Wise12,Pallottini14}.

Star formation follows a Schmidt-Kennicutt relation \citep{Kennicutt98}, dependent on the molecular hydrogen gas density, and assumes a Kroupa \citep{Kroupa01} initial mass function (IMF) for stellar particles. Stellar feedback mechanisms, including supernova explosions, winds, and radiation pressure, are modeled as described in \cite{Pallottini17}. The energy inputs and chemical yields, which depend on stellar age and metallicity, are computed using \code{Starburst99} \citep{starburst99} with Padova stellar tracks \citep{bertelli94}, covering a metallicity range of $Z_\star / Z_\odot = 0.02 - 1.0$. 

To investigate the properties and spatial distribution of dense neutral gas in the ISM and CGM of high-$z$ galaxies, we select a sample of $\sim 100$ central galaxies. 
The sample consists of central galaxies selected to uniformly span the halo mass range $M_h = 10^{9.5-11.5} M_{\odot}$ and the redshift range $z = 9.5$–6.0, in order to investigate the dependence of galaxy properties on both halo mass and redshift. These halos correspond to galaxies with UV magnitudes in the range $-22.5 \lesssim M_{\rm UV} \lesssim -18$, in the range of JWST spectroscopy. Our focus here is on generally understanding the nature and evolution of dense HI in and around galaxies, so we leave a more detailed comparison to observations, accounting for selection functions, to future work.
\code{serra} galaxies in these mass and redshift ranges have been shown to successfully reproduce several observed properties of \highz galaxies, such as the normalization and scatter of the mass metallicity relation \citep{Pallottini24}, the burstiness of the star formation histories \citep{Pallottini23, Gelli25} from JWST, as well as ISM and CGM gas tracers -- [C\,{\sc ii}]-$158\mu$m and [O\,{\sc iii}]-$88\mu$m emission observed by ALMA \citep{Pallottini2022, Kohandel24, Vallini2025}. For these reasons, we expect the simulations to reasonably reproduce the distributions and abundances of neutral gas in the ISM and CGM of \highz galaxies.

In this work, we take advantage of this diverse sample of \highz galaxies to investigate how the distribution of neutral hydrogen depends on their properties and redshift evolution. A key step involves deriving column densities $\NHI$ in a manner analogous to observations of galaxies with JWST.
To do this, for each galaxy, we (i) extract a $100$~kpc size box centered on each galaxy (to ensure that it includes the entire virial halo of the galaxies in the mass range considered), (ii) identify the location of peak of star formation by locating the center of mass of young stars ($<5$~Myr), corresponding to the brightest UV-emitting region typically targeted by JWST, and (iii) generate sightlines originating from this point that extend radially to the edge of the simulation box in $\sim 1000$ random directions. We then compute the column density along each sightline by integrating the neutral hydrogen density profile: $\NHI = \int_{\rm Sightline} n_{HI}(r)\, \mathrm{\rm d}r$, with $n_{HI}$ from \code{KROME}.

\begin{figure}
\centering
\includegraphics[width=0.49\textwidth]{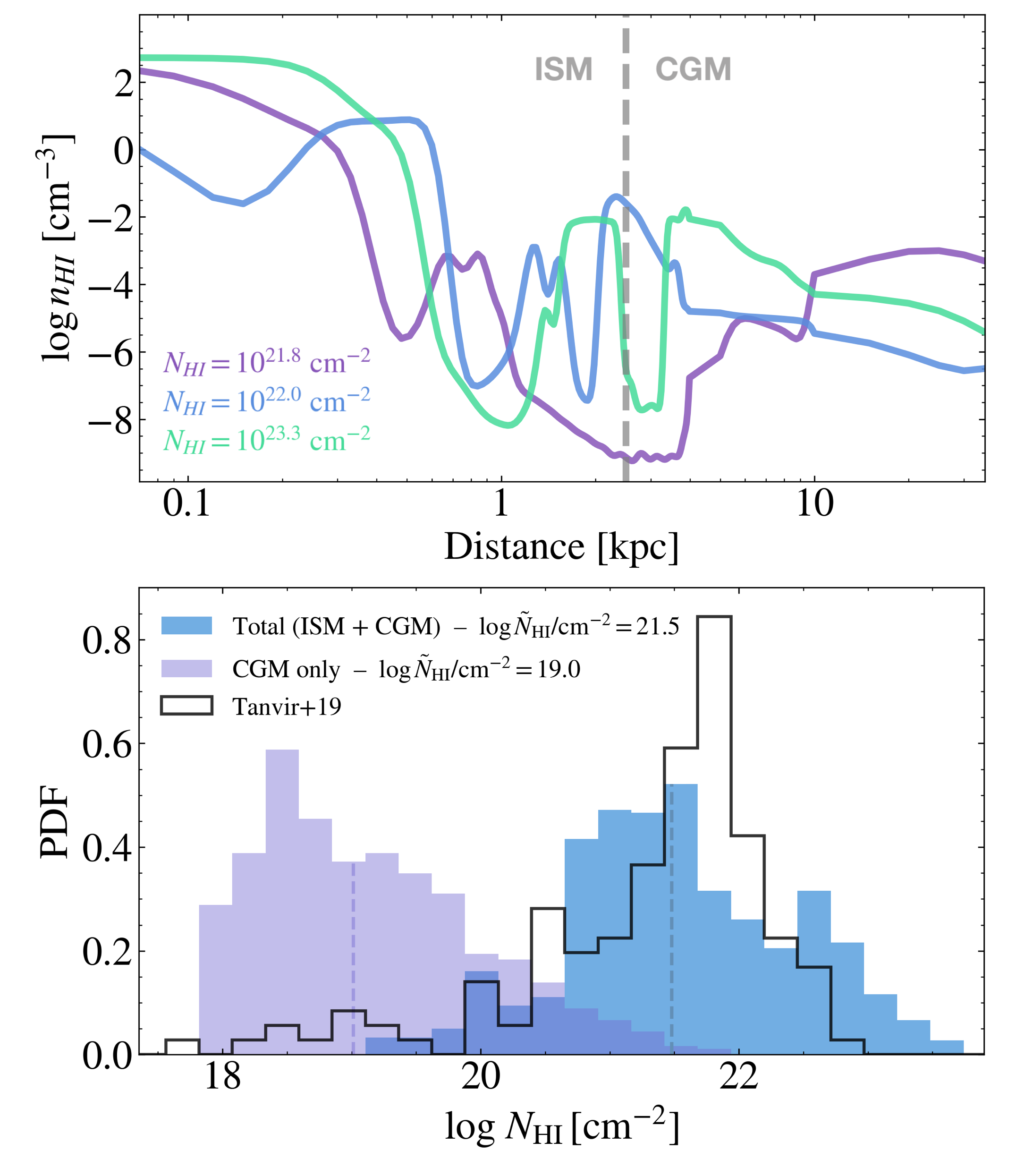}
\caption{\textit{Top}: HI density along three different sightlines from the centre of the galaxy \textit{Amana}. The resulting column density along each sightline targeting the same galaxy can thus vary significantly. \textit{Bottom}: Probability distributions of HI column densities along 3000 random sightlines towards the central galaxy (blue) and the same distribution only considering the contribution of the CGM to the column density (lilac) obtained by excluding the inner 2.5~kpc. The black histogram show the results of GRB DLA at $z\sim 2-6$ by \cite{Tanvir19}.
\label{fig:sightlines_pdf}}
\end{figure}

\section{Results} \label{sec:results}

Here, we first describe in depth a single representative galaxy's gas properties and spatial distribution (Sec.~\ref{subsec:one_galaxy}), and then expand the analysis to the full sample to explore mass (Sec.~\ref{subsec:properties}) and redshift (Sec.~\ref{subsec:z_evolution}) dependencies.

\subsection{Neutral gas distribution around a $z=7$ galaxy}\label{subsec:one_galaxy}

To get a first look at high-z galaxies' complex structure and gas distribution, in Fig.~\ref{fig:z8_galaxy} we show a typical $M_h \sim 10^{11}\msun$ halo at $z=7$, hosting a $M_\star \sim 10^{9.5}\msun$ star-forming galaxy in its center, called \textit{Amana}. The three panels show the neutral hydrogen column density, stellar surface density, and gas metallicity, all integrated along the $z$ direction, in a box of physical size of $\rm 100~kpc$.
Throughout this work, we refer to the ISM as all gas within 2.5~kpc of the central star-forming region, and to the CGM as all gas beyond this radius. The 2.5~kpc threshold is a conservative choice that ensures we include the entirety of the ISM, as the typical gas half-mass radius in high-redshift galaxies in the mass range considered is $\lesssim 1.5$~kpc \citep[e.g.,][]{Pallottini19, Pallottini2022, kohandel:2020}.

Fig.~\ref{fig:z8_galaxy} shows that the highest \hi\ gas concentration is reached in the inner 2.5~kpc, where the column density can reach up to $\NHI \sim10^{23} \rm cm^{-2}$ \citep[similarly to the simulated galaxy \textit{Alth{\ae}a}, see][in particular Fig. 2 therein]{Behrens2019}.
The insets zooming onto the ISM of \textit{Amana} show that most of the stars are located in the densest central regions, where winds from massive stars and supernovae efficiently enrich the medium with metals up to $Z_{\rm gas}>0.1\rm~Z_\odot$ \citep[see also][]{Pallottini24}. At the same time, mechanical feedback produces winds and shock-driven outflows, while metal-poor gas is continuously accreted from the surrounding medium. The interplay of these processes (mechanical and chemical feedback, frequent mergers, gas accretion) gives rise to bursty star formation \citep[][]{Pallottini23} and temporary quenching \citep{Gelli25} in early galaxies, which imprints a complex morphology to the neutral gas, distributed irregularly around the central disk and with column densities that can vary by up to $\sim5$ orders of magnitudes in the inner $\sim2.5$~kpc.

Moving outward from the innermost region, we see the neutral gas in the CGM is mostly being accreted onto the galaxy through multiple dense filaments, with $\NHI \gtrsim 10^{20} \rm cm^{-2}$. Within and in proximity to these filaments, the gas is seen to condense in numerous clumps reaching $\NHI \gtrsim 10^{22} \rm cm^{-2}$. With the exception of a few star-forming satellite galaxies (typically associated with metal enriched gas, see also \citealt{Gelli20}), the vast majority of these dense \hi\ clumps are not forming stars and are composed exclusively of pristine gas. 
They are present in large numbers around \highz galaxies, with a few tens of clumps typically residing within the virial radius of $\sim 20$~kpc for a $10^{11}\msun$ halo.
However, as we will show below, the probability of intersecting such a clump along the line-of-sight is low.

To assess the relative contributions of \hi\ gas at different distances in affecting observations of high-redshift sources and their \Lya absorption profiles, we analyze the spatial distribution of this intervening dense medium along narrow-beam sightlines targeting the central galaxy.
In the top panel of Fig.~\ref{fig:sightlines_pdf}, we show the radial profile of neutral hydrogen density for 3 independent sightlines targeting the central star forming region of \textit{Amana}, centered on the peak of star formation of the central galaxy. 
In the central $<100$\,pc regions, the density can reach up to $\gtrsim 10^2 \rm cm^{-3}$, corresponding to the dense star-forming molecular clouds in the galaxy's ISM. However, already within $\lesssim 1$~kpc the profiles can significantly differ from one another. Depending on whether a sightline intercepts a low-density channel cleared by supernovae feedback, inflowing gas accreting through a filament or other star-forming regions, the density can vary by up to $\sim7$ dex.
We note, numerical relaxation may also displace star particles from their birth clouds, which could also contribute to a wide range of HI densities in the inner regions.
All profiles tend to drop to $< 10^{-3} \rm cm^{-3}$ as we move towards the outer CGM and beyond the virial radius, but still with strong variation between different sightlines.
As a consequence, the inferred neutral hydrogen column densities show a strong sightline dependency and can vary by several orders of magnitude for the same galaxy depending on the observing angle.

To quantify the column density sightline variability, we draw 3000 random line-of-sight directions from the central galaxy and plot the probability distribution of $\NHI$ in the bottom panel of Fig.~\ref{fig:sightlines_pdf} (blue histogram).
The distribution is broad, varying from $10^{19}$ to $10^{23.5}\rm cm^{-2}$, and with median $10^{21.5}\rm cm^{-2}$. 
The width of the distribution, spanning over $\gtrsim3$ orders of magnitude, highlights the strong sightline dependence of the observed column densities. This may be driven by the complex gas morphology in both the ISM, governed by star formation–related processes (stellar feedback driving outflows, frequent merger events), and the CGM, characterized by filamentary and clumpy structures.

To understand where the gas driving these column densities is predominantly located, we separate the contributions of the ISM and CGM to the inferred column densities. We do this by recalculating \NHI for the same 3000 sightlines but excluding the inner $2.5$~kpc, effectively isolating the role of the CGM, including potential intervening dense clumps and satellites.
The obtained PDF (shown in lilac) is still wide, highlighting how the CGM itself can cause variations of $>4$ orders of magnitude in different sightlines' column densities. However, the distribution is shifted towards much lower values of $\NHI$, ranging from $10^{17.5}$ to $10^{21.5}\rm cm^{-2}$, and median at $10^{18.4}\rm cm^{-2}$. The probability of a sightline encountering a dense clump with $\NHI>10^{21}\rm cm^{-2}$ in the CGM is $\sim 2 \%$ for this galaxy, compared to $\sim 70 \%$ when also including the ISM.
Comparing the \NHI distributions for the CGM only versus CGM + ISM clearly shows that dense neutral gas in the inner ISM, inevitably intercepted by any sightline targeting star-forming regions, represents the dominant component of the absorbing gas and is expected to be the main driver of high column densities.

Interestingly, we find our example \hi\ column density distribution is very similar to that derived from observations of $z\sim2-6$ GRB afterglows \citep{Tanvir19}. 
This suggests that \highz galaxy DLAs probe similar star-forming regions as lower-$z$ GRB-DLAs, which originate in the inner ISM of galaxies from the deaths of massive stars. Thus, integrated high-$z$ galaxy sightlines appear to trace, on average, gas distributions similar to those revealed by GRB-DLAs.
We note the \NHI distribution inferred for \textit{Amana} shows a higher fraction of sightlines with $\NHI \simgt 10^{22.5}$\,cm$^{-2}$ compared to the GRB-DLA distribution, which could be linked to a observational bias as GRB detections will likely avoid the very high-column density and metal rich systems because of the significant reddening \citep[e.g.][]{Kruhler2013}. \textit{Amana} also does
not show the tail to low column densities which would allow Lyman continuum photons to escape (only gas $\NHI < 2 \times 10^{17}$\,cm$^{-2}$ is optically thin to Lyman continuum). Other examples in our sample do show lower column density sightlines, though they are rare. 
These differences between the simulated and inferred GRB-DLA distribuion may be related to the resolution limit of our simulations that cannot spatially resolve HII regions \citep{Pallottini19, decataldo:2019} and therefore possibly underestimate the abundance of low-density channels and overestimate high-density regions. We discuss this further in Section~\ref{subsec:disc_sightline}.

\begin{figure}[t!]
\centering
\includegraphics[width=0.5\textwidth]{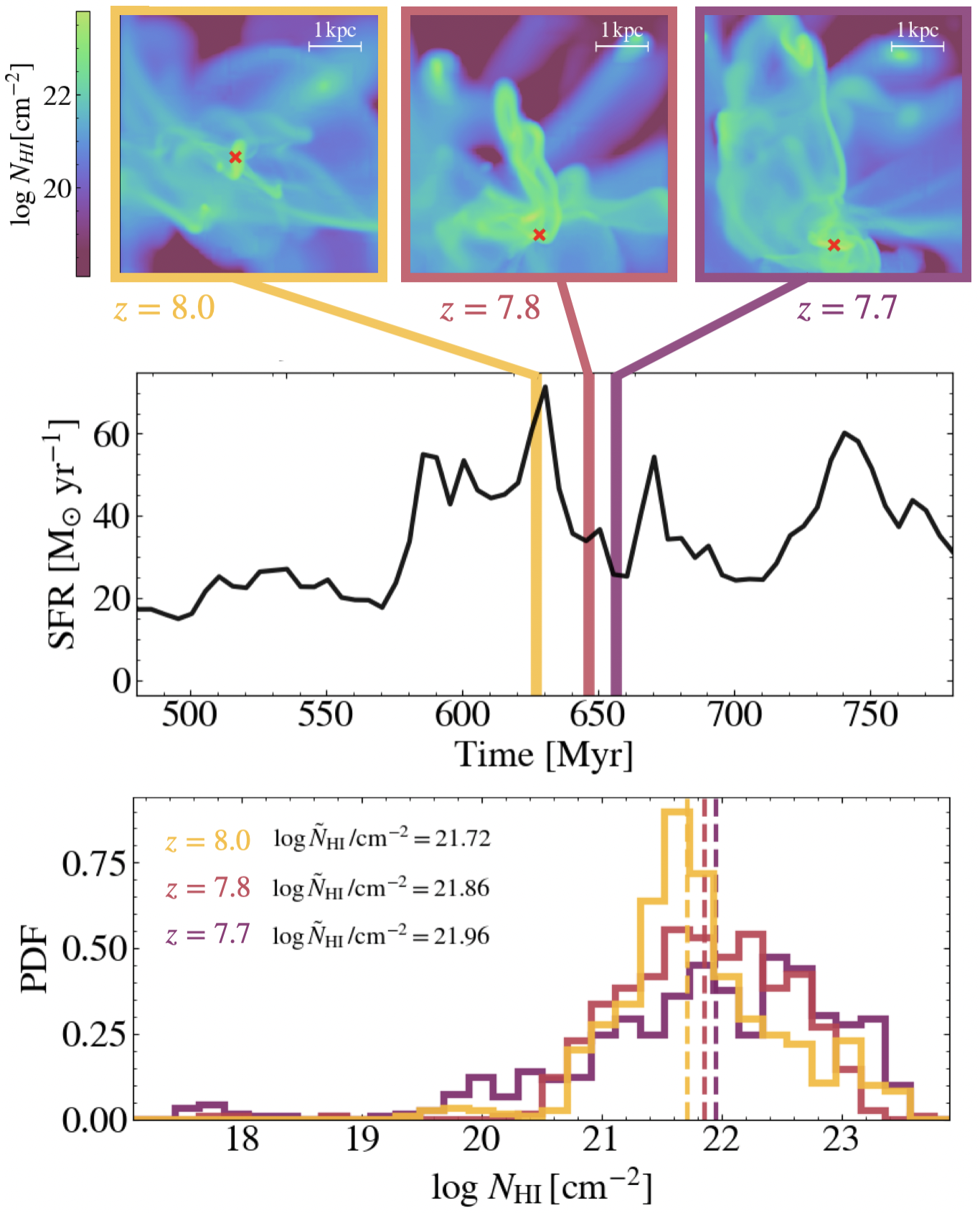}
\caption{Star formation history of a typical massive galaxy in \code{serra} (\textit{centre}). For three specific timesteps during its evolution, during and following a strong burst of star formation, we show the ISM column density maps over 5~kpc size boxes (\textit{top}) and column density distributions (\textit{bottom}). The red crosses in the maps pinpoint the location of the peak of star formation.
\label{fig:bursty_sfh}}
\end{figure}

\begin{figure*}[t!]
\centering
\includegraphics[width=\textwidth]{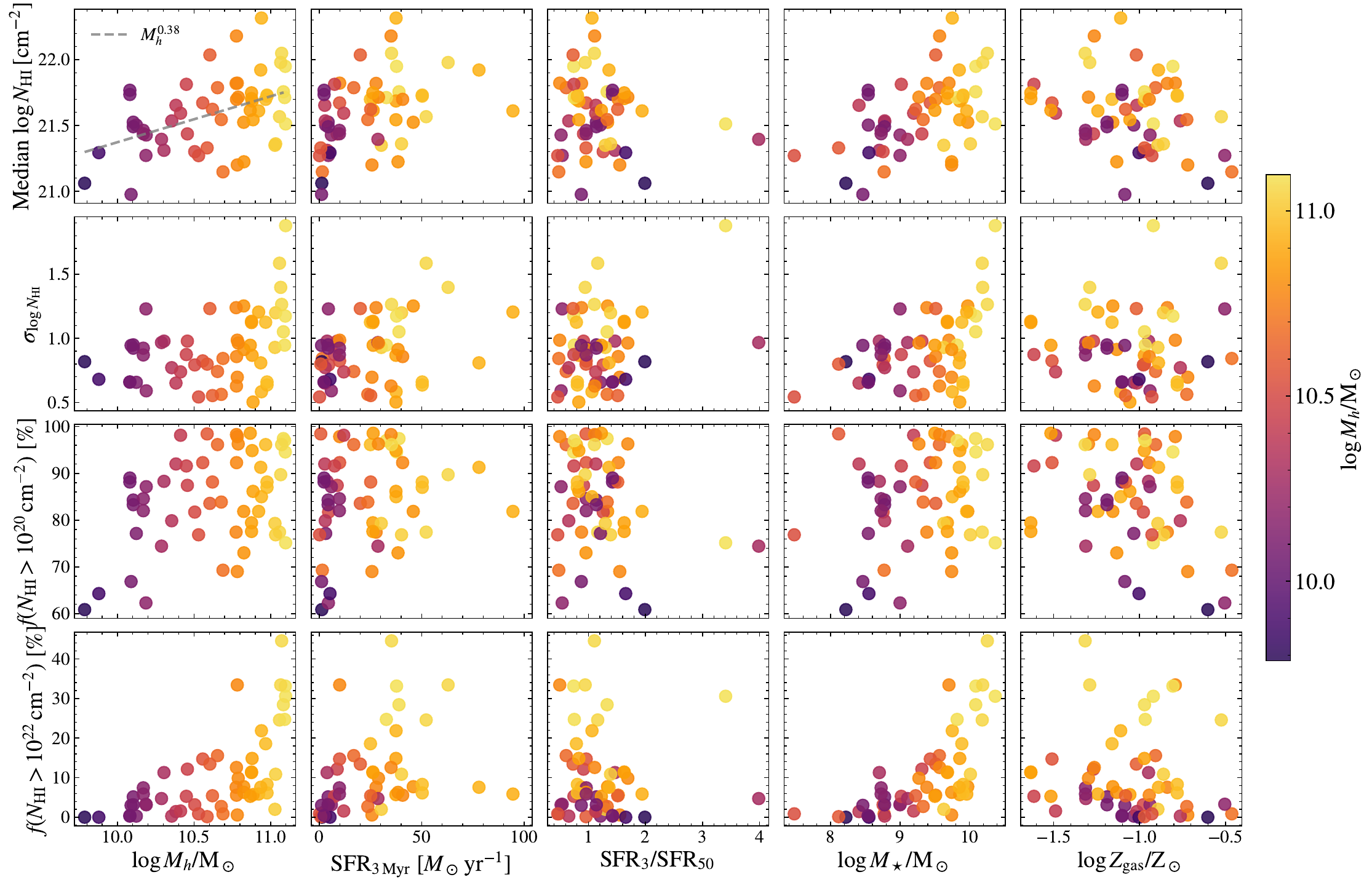}
\caption{Neutral hydrogen column density metrics for each galaxy in the sample, as a function of key physical properties. Each row corresponds to a different quantity derived from the $\NHI$ distributions along random sightlines: median, standard deviation of $\log(\NHI/\mathrm{cm}^{-2})$, and fraction of sightlines with $\NHI > 10^{20},\mathrm{cm^{-2}}$ and $\NHI > 10^{22},\mathrm{cm^{-2}}$. Each column shows results versus (from left to right): halo mass, star formation rate (averaged over the last 3 Myr), burstiness ratio ($\rm SFR_3/SFR_{50}$), stellar mass, and ISM metallicity. Points are color-coded by halo mass. A clear trend with halo mass is visible for all \NHI metrics, while burstiness and metallicity show no significant correlation with $\NHI$ properties. The dashed line in the first panel shows the result of the power-law fit of $\NHI\propto M_h^{0.38}$.
The large intrinsic scatter present in all the relations highlights the complex and inhomogeneous ISM structure within individual galaxies.
\label{fig:scatter_grid}}
\end{figure*}

We have seen that the large sightline-to-sightline variation in the inferred column densities for a single galaxy arises from the complex neutral gas morphology in its ISM. This complexity is driven by the highly bursty star formation of early galaxies, where frequent merger events and efficient inflow of gas induce starbursts, while intense stellar feedback drives powerful outflows.
To understand if and how the column density distributions are linked to bursty star formation, we examine how they evolve during a starburst cycle of another typical massive galaxy in \code{serra}. In Fig.~\ref{fig:bursty_sfh} we show its star formation history, and corresponding ISM \NHI\ maps at three key stages during its evolution: (i) at the peak of the strongest starburst, (ii) $\sim 20$~Myr later, when supernova feedback has substantially suppressed star formation, and (iii) $\sim 30$~Myr after the peak, during the accretion phase just before the next burst. The maps reveal a chaotic ISM that rapidly evolves and changes over just tens of Myr.
The corresponding \NHI\ histograms at these three stages, shown in the bottom panel, reveal that, at the starburst peak, the distribution is peaked at $\log \NHI/\rm cm^{-2} = 21.7 $. Following the burst, the distribution broadens slightly as feedback clears low-density sightlines while others intersect compressed ejected gas and newly inflowing dense reservoirs that will form stars within the next 5–10~Myr. The fraction of sightlines with $\log \NHI/\rm cm^{-2} < 21$ increases from 8\% during the bursts to 11\% and 19\% for the following timesteps, reflecting the progressive opening of low-\NHI\ channels after the burst.
However, the changes in the median column density and PDF width remain very modest ($\lesssim 0.2$~dex change in the median $\log$\NHI) indicating that, despite feedback and inflows inducing significant morphological variations, the overall column-density distribution remains remarkably unchanged. It might be possible that with higher resolution resolving molecular cloud scale and evolution, a further dependence of the mean column density could appear on shorter timescales on which GMC evolve ($<10$\,Myr).

We next extend our analysis to a larger sample of galaxies across the epoch of reionization to determine whether these findings hold for different galaxy properties and redshifts.

\subsection{What is the origin of high-z galaxy-DLAs?}\label{subsec:properties}

The analysis of \textit{Amana} revealed that the local ISM is the dominant contributor to high neutral hydrogen column densities, with its complex morphology leading to a high sightline-to-sightline variation. We now proceed to assess whether this holds across a broader sample, and how the inferred column densities depend on galaxies' properties and redshift. We do this by studying $\sim100$ \code{serra} galaxies spanning redshifts $z = 6 - 9.5$ and halo masses in the range $M_h = 10^{9.5} - 10^{11.5} M_{\odot}$.  

For every galaxy, we evaluate the distribution of $\NHI$ along multiple sightlines, following the same approach as in the previous section. In Fig.~\ref{fig:scatter_grid}, we present four summary statistics for the \NHI distribution for each galaxy in our sample: the (1) median and (2) standard deviation of the $\NHI$ distribution, as well as the fraction of sightlines with column densities exceeding thresholds of (3) $10^{20}$ and (4) $10^{22}\,\rm{cm^{-2}}$. 
We show these summary statistics as a function of key physical parameters shown in each column: halo mass, star formation rate (averaged over the last 3 Myr, index of the instantaneous SFR as inferred from tracers like Balmer emission lines), the burstiness ratio $\rm SFR_3/SFR_{50}$ (index of recent starburst activity; see \citealt{Gelli25}), stellar mass and ISM metallicity (defined as the mass-weighted gas metallicity in the galaxy's ISM). Points are color-coded by halo mass.
\begin{figure*}[t!]
\centering
\includegraphics[width=\textwidth]{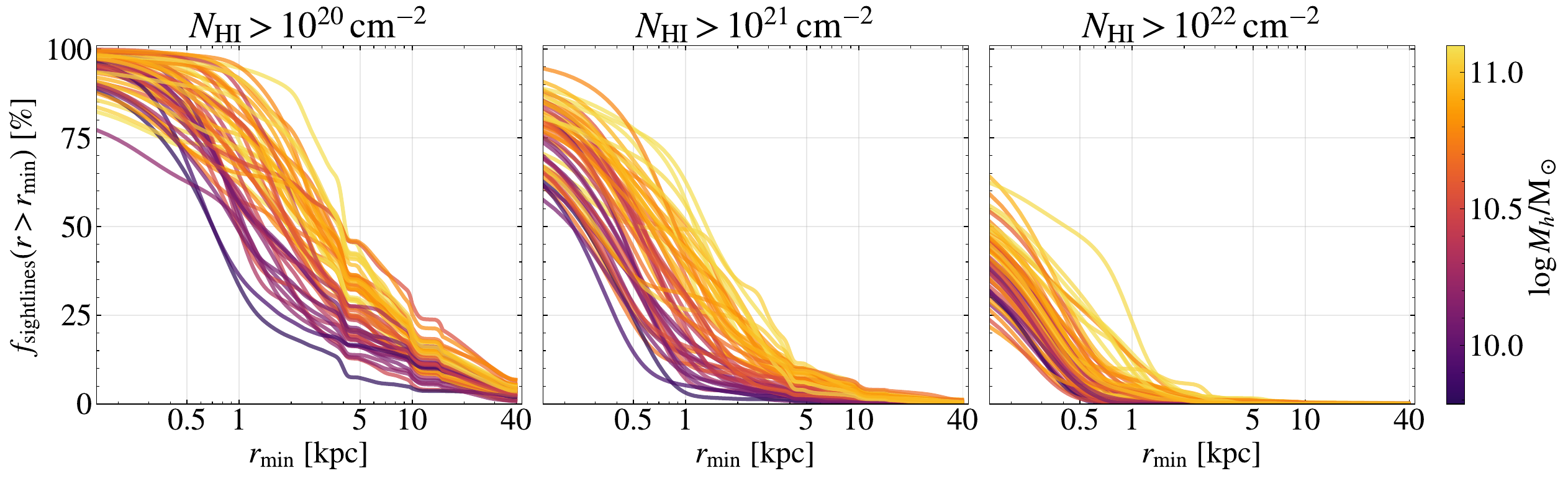}
\caption{Fraction of sightlines with column density above thresholds $\rm 10^{20} cm^{-2}, 10^{21} cm^{-2}, 10^{22} cm^{-2}$ for all galaxies in our sample from \code{serra}, color coded with their halo mass. The figures illustrate a clear radial decline, with the inner ISM contributing most significantly to the probability of encountering high column densities. A strong dependence on halo mass is also evident across all thresholds.
\label{fig:fractions_vs_r}}
\end{figure*}

We find the median column densities and the probability of encountering high \NHI increase with halo mass.
Assuming a power-law \hi\ density profile out to the virial radius \citep[see e.g.,][]{Stern2021, pallottini:2014b}, we would expect \NHI~ to scale roughly with virial radius, i.e. $\NHI \sim M_h^{1/3}$.
Fitting the trend between median \NHI and halo mass in our sample with a power law we find a scaling very similar to this expectation: $\NHI = 10^{21.73} (\frac{M_h}{10^{11}\rm M_\odot})^{0.38} \, \rm cm^{-2}$.
The increasing trend with halo mass supports the picture in which more massive halos, with their deeper potential wells, are more efficient at accreting and retaining gas from the cosmic web; at the same time, their larger virial radii naturally increase the path length through which neutral gas is integrated, further boosting the column densities.
The slightly steeper trend we recover compared to our simple estimate may imply higher \hi\ column densities in massive halos due to their more clustered environments, and lower column densities in low mass halos which may be more efficient at ionizing/removing dense HI due to feedback.

On the other hand, the remaining physical properties do not exhibit clear correlations with the \NHI distribution, with the color-coding indicating halo mass dominates the \NHI distribution. 
We find a mild increase of the median \NHI with $\mathrm{SFR}_3$ and $M_\star$, which is likely driven by the underlying positive correlation between these quantities and halo mass, but with significant scatter as SFH is bursty and highly time-variable at fixed halo mass. We also find a slight anti-correlation, though with large scatter, between median \NHI and ISM metallicity. This may arise from more massive halos halos — on average with higher \NHI — experiencing stronger dilution from the efficient accretion of metal-poor gas.

\begin{figure*}[t]
\centering
\includegraphics[width=\textwidth]{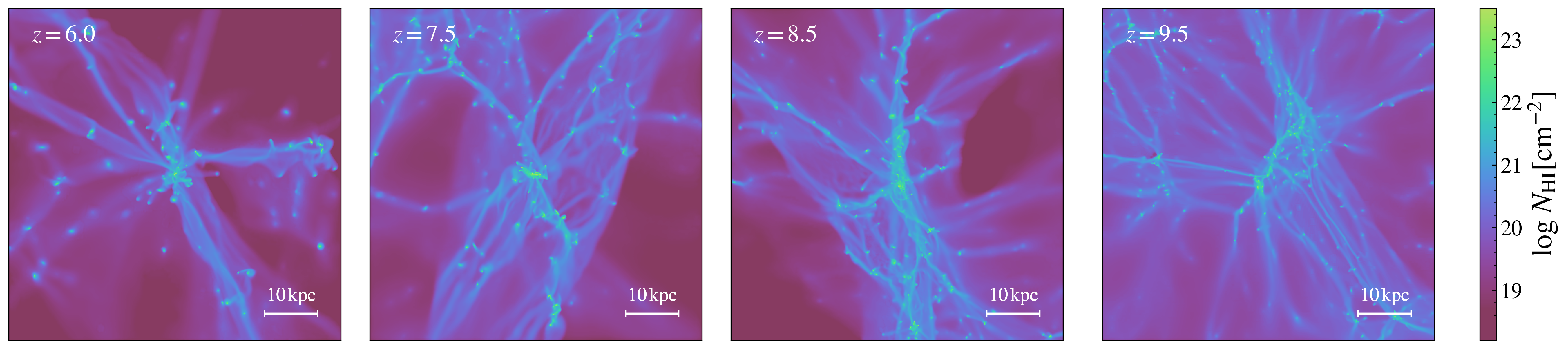}
\caption{Neutral hydrogen column density maps of central halos of similar mass $M_h=10^{11}\msun$ at different redshifts, over a box of 100~kpc.}
\label{fig:zevol_maps}
\end{figure*}

While halo mass appears to be the primary driver of the \NHI distribution, reflecting a galaxy's larger virial radius and ability to efficiently accrete gas from its surroundings, the standard deviation of each galaxy’s $\log \NHI/\rm cm^{-2}$ distribution always remains above $\sigma_{\log \NHI} > 0.5$. This indicates that sightline-to-sightline variations within individual galaxies are expected to produce significant scatter in the observed column densities. We find no clear trend of $\sigma_{\log \NHI}$ with galaxy properties, though we see that the highest values of $\sigma_{\log \NHI}>1.5$ are reached only for the most massive halos in our sample ($M_h > 10^{11}\,M_\odot$), which we find reflects their more complex ISM morphologies and an increased abundance of filaments and dense HI clumps in CGM, as well as the higher abundance of possible intervening nearby satellites.

To quantify the relative importance of dense neutral gas located at different distances from the emitting galaxy, we vary the lower integration limit of $r_{\rm min}$ for each sightline (where $\NHI = \int_{r_{\rm min}}^{50 \rm kpc} n_{\rm HI} dr$) from the peak of the star-forming region outward. In Fig.~\ref{fig:fractions_vs_r}, we show the fraction of sightlines with $\NHI > 10^{20,\, 21, \, 22} \,\rm cm^{-2}$ as a function of $r_{\rm min}$ for all galaxies, color-coded by their halo mass. This can be interpreted as a radial covering fraction, or as the probability of a galaxy intersecting an absorption system above a certain column density threshold at a certain radius.

The figure highlights multiple key trends. Across all galaxies in our sample, the central $\lesssim 1$ kpc region is the primary contributor to high $\NHI$ values. The decline in the fraction of sightlines intersecting high \hi\ columns with increasing $r_{\rm min}$ implies that the probability of encountering a high column density sightline is highest in the ISM and declines towards the CGM. In Figure~\ref{fig:z8_galaxy} we have seen that tens of self-shielded, dense \hi\ clumps with $\NHI > 10^{22}\,\mathrm{cm}^{-2}$ are present in the CGM surrounding high-redshift galaxies. Here the radial trend reveals that the likelihood of such structures of intersecting a random line of sight to the central galaxy is extremely low, always below $\lesssim 2\%$. As a result, their contribution to the average incidence of strong absorption is negligible compared to the dense ISM in star-forming regions.

The fraction of sightlines with high $\NHI$ decreases with increasing column density threshold, i.e. going from the left to the right panels in Fig.~\ref{fig:fractions_vs_r}. This reflects the increasing rarity of extremely dense neutral gas sightlines, with $\NHI > 10^{22} \,\mathrm{cm}^{-2}$ being significantly less common than $\NHI > 10^{20} \,\mathrm{cm}^{-2}$. 
We find the majority of our sample have a $>70\%$ probability of producing sightlines with $\NHI > 10^{20} \,\mathrm{cm}^{-2}$ and a $\gtrsim 60\%$ ($\gtrsim 30\%$) probability of reaching $\NHI > 10^{21} (10^{22}) \,\mathrm{cm}^{-2}$. A proper comparison with observations would require detailed matching with observational selections which is beyond the scope of this work, but we note these fractions are comparable to the findings of \citet{Mason25}, who investigated a sample of $\sim100$ high S/N JWST spectra in a similar UV magnitude range, accounting for \Lya emission and IGM attenuation, and found $\sim18\pm5\%$ of $z>5$ galaxies show $\NHI \simgt 10^{22} \,\mathrm{cm}^{-2}$. The results are also consistent with \cite{Heintz25}, who found that $\gtrsim 60\%$ of galaxies in the JWST-PRIMAL sample at $z>5$ have $\NHI \simgt 10^{21} \,\mathrm{cm}^{-2}$.

The color coding in Fig.~\ref{fig:fractions_vs_r} reveals a clear halo mass dependence: more massive galaxies have a greater probability of exhibiting high $\NHI$ sightlines.
The mass dependence is particularly evident also in the CGM component, where cold gas accretion through filaments is primarily regulated by the depth of the gravitational potential.

When rescaled by virial radius, we find the profiles of all halos converge more closely, indicating self-similarity in the gas density distribution \citep[e.g.,][]{pallottini:2014b,Rahmati15,Stern2021}.

The observed mass dependence has important implications for observations. Since lower mass halos are more abundant in the early universe, we expect the \hi\ covering fractions of the observed galaxy population to decrease on average. This trend could help interpret recent results from low-ionization absorption lines (LIS) studies at $z\sim3-7$ by \citet{Pahl2023} and \citet{Glazer25}, who derived decreasing \hi\ covering fractions with increasing redshift, which may reflect the lower average halo masses at earlier times. As we discuss in in Sec.~\ref{subsec:disc_DLAs}, simple clustering analyses may be able to test this picture, as more massive halos are expected to have a higher number of close companions. 

We note, throughout the paper, all $\NHI$ estimates are derived using gas within $\lesssim 50$~kpc of galaxies. However, intervening dense gas located in clumps and galaxies at larger distances could also contribute to the observed column densities as proximate absorbers \citep[e.g.,][]{Hennawi2007,Christensen2023,Davies2023}. We tested the potential impact of such absorbers by extrapolating the $\NHI$ distribution into the regions outside the CGM ($r > 40$~kpc) over a redshift range of $\Delta z \sim 0.2$, where absorbers could produce damping wing absorption around rest-frame \Lya wavelengths. We find fewer than $10 \%$ of sightlines exceed $\NHI > 10^{21} \rm{cm}^{-2}$. This represents a very conservative upper limit, as the estimate is based on extrapolation of very dense environments where we selected our halos. We therefore expect that high column density absorption from intervening gas at larger distances is very uncommon and does not affect our conclusions.

In the next section, we examine how the neutral gas distribution and column densities around the halos in our sample evolve with redshift.

\subsection{Redshift evolution}\label{subsec:z_evolution}

Observationally, most studies suggest minimal redshift evolution in \hi\ column density distributions \citep[e.g.,][]{Reddy2016, Umeda2024, Mason25, Heintz25}, but a potential decrease in \hi\ covering fractions at higher redshifts \citep[][]{Pahl2019,Glazer25}. However, the extremely high levels of scatter in the inferred \NHI~ distributions from large samples of galaxies makes it challenging to identify and interpret these trends. In principle, higher redshifts halos are expected to have higher gas densities, which could favor increased column densities. Conversely, if massive DM halos are the primary contributors to large $\NHI$, their declining abundance at higher redshifts may lead to lower average column densities.  

To investigate this evolution, we analyze how the expected \hi\ column densities in \code{serra} galaxies change from $z \sim 9.5$ to $6.5$. In Fig.~\ref{fig:zevol_maps}, we show galaxies with similar halo masses of $M_h \sim 10^{11} \msun$ at different redshifts. From the maps we see no significant redshift evolution in the overall distribution of \hi\ within $\lesssim 50$~kpc, with the central galaxy residing at the intersection of at least multiple filaments and reaching high column densities in the central ISM ($\NHI > 10^{21} \rm cm^{-2}$). Dense neutral gas in the CGM is more abundant at higher redshifts, while at lower redshifts the CGM is instead characterized by more extended, lower-density regions.

\begin{figure}[t]
\centering
\includegraphics[width=0.4\textwidth]{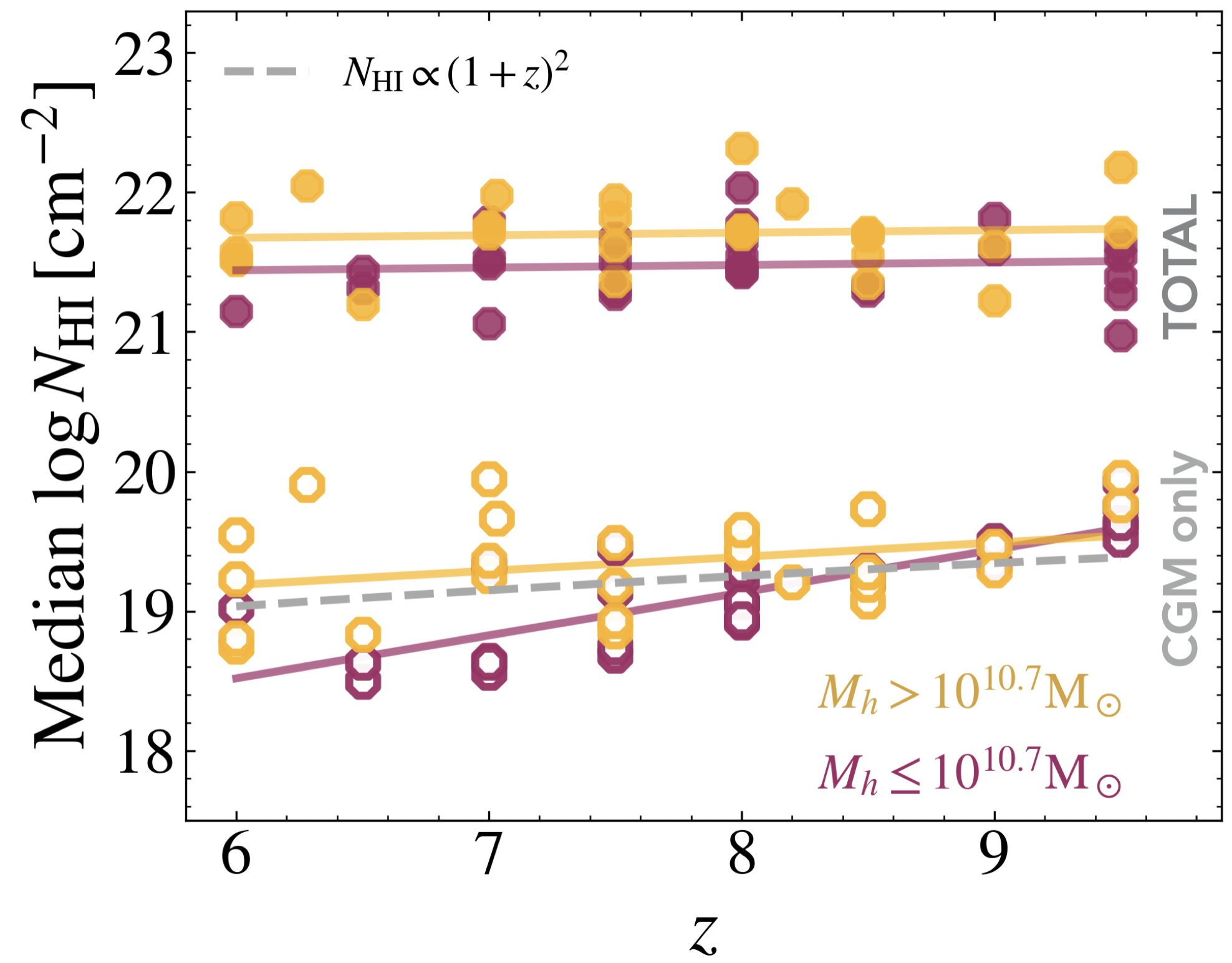}
\caption{Redshift evolution of the medians of the column densities distribution of \code{serra} galaxies, for the total (ISM + CGM) sightlines and CGM only ($r>2.5$~kpc).}
\label{fig:NHI_redshift}
\end{figure}

To quantify the evolution of ISM and CGM column densities, Figure~\ref{fig:NHI_redshift} shows the median of the $\NHI$ probability distribution function as a function of redshift for the entire galaxy sample also analysed in the previous sections. The filled and empty points identify the two components, total (ISM+CGM) and CGM only respectively (defined as in Figure~\ref{fig:sightlines_pdf}).
We divide the galaxies in two halo mass bins to highlight the mass dependence, and the solid line shows a linear fit of the evolution in each subsample to guide the eye.

At all epochs, the typical total median column densities span $\NHI \sim 10^{21.0}-10^{22.5}\rm cm^{-2}$, with no significant redshift evolution -- only a mild increase scaling as $(1+z)^{0.3}$.
The color coding once again reveals that more massive galaxies tend to have higher median $\NHI$, indicating that halo mass dependence dominates over redshift evolution at fixed mass. The CGM-only column densities are characterised by much lower values and similar scatter $\NHI \sim 10^{18.5}-10^{20.0}\rm cm^{-2}$. As suggested by the maps, the median $\NHI$ through the CGM is subject to a more pronounced evolution, with higher column densities at higher redshifts. The trend roughly follows the cosmological mean hydrogen density with column density scaling as $N_{\rm H} \propto (1+z)^2$, shown by the gray dashed line. We discuss this redshift evolution further in Section~\ref{subsec:disc_CGM}.

We also emphasize that, as seen in Sec.~\ref{subsec:properties}, the sightline-to-sightline variation leads to a scatter of $\sigma_{\log \NHI} > 0.5$ for a single galaxy, dominating over redshift evolution. Therefore, we do not predict significant changes in the \NHI distributions from the end of reionization up to $z\sim 9$.

\section{Discussion} \label{sec:discussion}

Our results indicate that galaxy-DLAs at $z>6$ primarily arise in the dense gas in the ISM of \highz galaxies, with large sightline variation towards star-forming regions due to inhomogeneous gas distributions in the ISM and CGM. In this section we discuss the implications of this broad sightline variance for interpreting high-redshift spectra in the context of reionization (Section~\ref{subsec:disc_sightline}); the nature of extremely strong DLAs and predictions for their environments and metallicity (Section~\ref{subsec:disc_DLAs}); and discuss the role of evolution in the ISM and CGM during the epoch of reionization (Section~\ref{subsec:disc_CGM}).

\subsection{The implications of sightline variance in HI distributions} \label{subsec:disc_sightline}

In Section~\ref{subsec:one_galaxy} we showed that sightlines towards star-forming regions at $z\sim6-9.5$ exhibit a broad distribution of \hi\ column densities due to inhomogeneous gas distributions in the ISM and CGM. We find the highest \NHI absorbers ($>10^{22}$\,cm$^{-2}$) arise in the inner ISM ($<1$kpc). Our results suggest that the halo mass is the primary driver of the \hi\ gas distribution and derived column density distribution around galaxies.

These findings are qualitatively consistent with previous studies using cosmological simulations to analyze \hi\ around galaxies at lower redshifts ($z<6$).
Many previous works have focused on interpreting absorption in background quasar spectra. For example, \cite{Faucher-Giguere15}, and more recently \cite{Tortora24}, found a steep decrease in the abundance of Lyman limit systems ($\NHI \gtrsim 10^{17}$\,cm$^{-2}$) with increasing radius from the center of galaxies in the \code{FIRE} simulations at $z\sim 0-6$, and a higher covering fraction of \hi\ for more massive halos (which we will discuss more below), consistent with our results. Similar qualitative trends are also found in the \code{EAGLE} simulations \citep{Rahmati14, Rahmati15}.
Our results are also consistent with previous work examining sightlines directly towards star-forming regions. 
\cite{Pontzen10} investigated the origin of GRB afterglow \hi\ absorption at $2<z<4$, simulating sightlines emerging from galaxies, similar to our approach.  Their recovered \NHI distribution is very similar to ours, with a predicted median column density of $\NHI=10^{21.5}\rm cm^{-2}$, comparable with the median we find in \code{serra} for $M_h\sim 10^{9.5}-10^{11.5}\msun$ halos.
Based on their recovered \NHI distribution, they inferred GRB-DLAs arise within $\lesssim 1$~kpc from the center of the halo. 
High resolution simulations of gas around star-forming regions focused on understanding the escape of Lyman continuum and Lyman-$\alpha$ photons \citep[e.g.,][]{Ma2020,Kimm2019,Kimm2025} have also found similarly broad \NHI distributions, peaked around $\NHI=10^{20-22}\rm cm^{-2}$, arising from the interplay of feedback in dense star-forming clouds.
Large sightline variance in the \hi\ distribution in the ISM is also predicted to play a major role in driving the observed variance in \Lya emission line strengths and shapes \citep[e.g.,][]{Smith2019,Blaizot2023,Kimm2025}, consistent with an observed anti-correlation between \Lya equivalent width and the HI covering fraction derived by \citet{Reddy2022}.

Nevertheless we find a high covering fraction, $\gtrsim80-90\%$ of DLAs ($\NHI > 10^{20.3}$\,cm$^{-2}$) towards our halos, which increases with increasing halo mass. This is consistent with observations at $z\sim2-3$ finding a median $\NHI \approx 10^{20.3}$\,cm$^{-2}$ and covering fraction $\approx90$\% \citep{Reddy2016}, as well as recent JWST observations implying median $\NHI \approx 10^{20-21}$\,cm$^{-2}$ in $z\sim5-14$ galaxies \citep{Mason25}. Such a high covering factor of high column density neutral gas is also consistent with low inferred Lyman continuum escape fraction at all redshifts \citep[with average escape fractions $\lesssim 10$\% in star-forming galaxies, e.g.,][]{Steidel2016,Begley2022,Kreilgaard2024}.

In Figure~\ref{fig:fractions_vs_r} we have shown that the covering fraction of dense \hi\ increases with increasing halo mass. Observations of low ionization interstellar (LIS) metal absorption lines have demonstrated a decline in line equivalent width at fixed $M_\mathrm{UV}$ with increasing redshift, which may be interpreted as a decrease in the covering fraction of neutral gas \citep[e.g,.][]{Jones2013,Du2018b,Pahl2019,Glazer25}. Due to rising baryon accretion rates, galaxies at fixed $M_\mathrm{UV}$ are expected to trace lower mass halos with increasing redshifts \citep[e.g.,][]{Mason2015,Behroozi2015a}. Our results thus imply that a decrease in covering fraction may be expected at higher redshifts: as we observe increasingly lower mass halos which are less able to retain high density \hi.
We note that because we focus on zoom-in simulations of central galaxies, our sample is likely biased toward relatively massive systems ($M_\ast \gtrsim 10^{9.5}\,\mathrm{M}_\odot$) and their satellites. Future work could extend this analysis to include more isolated, lower mass, galaxies, to better capture the impact of local environment on HI gas distributions.

We also note the spatial resolution of our simulations, limited to $\sim$30~pc (comparable to giant molecular clouds), will not fully capture the complex, multi-phase, substructure of the ISM. 
For example, observations of \Lya emission with $\sim100-200$\,km/s red peak velocity offsets (implying $\NHI \sim 10^{19-20}$\,cm$^{-2}$ coincident with DLA absorption troughs implying $\NHI \sim 10^{20-22}$\,cm$^{-2}$ suggest a complex, multi-phase ISM structure \citep[e.g.,][]{Hu23} which is not resolved by galaxy-scale hydrodynamic simulations \citep[see discussion by][]{Gronke2017b}. It is difficult for \Lya to escape when molecular clouds are unresolved in simulations \citep{Behrens2019}.
In particular, without sufficient resolution, low density channels carved by stellar feedback, such as ionizing radiation winds and supernovae, are smoothed over \citep{decataldo:2020, dimascia:2025}, thus potentially underestimating the presence of low-column-density sightlines and overestimating the abundance of high column density sightlines. As a result, while the peak of the $\NHI$ distribution is robustly set by the average ISM densities, we may be missing a low-$\NHI$ tail -- we find no sightlines through the ISM with $\NHI < 10^{17} \rm cm^{-2}$. This component will also be important for enabling the escape of Lyman Continuum photons.

The \NHI distribution is also sensitive to feedback prescriptions \citep{Rahmati15,Faucher-Giguere15,Bennett2020}, with particular recent attention on the impact of cosmic ray (CR) feedback \citep[e.g.,][]{Girichidis2018}. Our simulations include CR feedback only via ionization \citep[][]{Pallottini17}, however, cosmic rays can provide non-thermal pressure, which can drive smooth, slow gas outflows, resulting in a more uniform and extended HI distribution in the ISM and CGM \citep{Girichidis2018}. \citet{Gronke2018a} and \citet{Kimm2025} have shown that this reduces the number of very low and high \NHI sightlines, producing slightly narrower \NHI distributions, with $\sim1\,$dex lower median \NHI compared to simulations without CR-driven outflows, and that this may also help produce more realistic \Lya emission properties.
Overall, pushing toward more realistic, higher-resolution simulations will be important for understanding more complex ISM structures on sub-parsec scales. However, because of the expected increase of density, is difficult to properly resolve HII regions even in simulations focusing their computational power on single molecular clouds \citep{decataldo:2019,decataldo:2020}, so progress will likely require a multi-scale approach. While there is uncertainty in the exact form of the \NHI distribution, our conclusion that it is broad remains unchanged.

A broad \NHI distribution has important implications for interpreting galaxy damping wings at $z>5$. It implies that individual galaxies may have a significant diversity in damping wing profiles due to the combination of inhomogeneous \hi\ distributions in the ISM, CGM and IGM, motivating dedicated fits to individual galaxy spectra which can marginalise over the expected variance.
We find minimal evolution in the local \NHI distribution with redshift at fixed halo mass as we discuss in Section~\ref{subsec:disc_CGM}. This suggests it may be possible to establish typical \NHI distributions for star-forming galaxies at the end of reionization, which can be used as priors for \NHI at higher redshifts when the IGM damping wing becomes significant, similar to the approaches used in interpreting the decline in \Lya emission at $z\simgt6$ \citep[e.g.,][]{Schenker2014,Treu2013,Mason2018,Tang2024,Tang2024c,Nakane24}.
Future deep moderate resolution spectroscopy ($R\gtrsim1000$) with JWST and ELT will enable a better separation of IGM and local \hi\ damping wings (see Section~\ref{subsec:disc_DLAs}), enabling this scenario to be tested more robustly in observations.
Furthermore, we note the sightline variance in \NHI implies conversions from \NHI to global \hi\ properties (e.g. gas mass) for individual galaxies may be subject to large systematic scatter, and may be better recovered with sample averages.

Finally, we note that JWST spectroscopy collects light through an aperture, integrating emission from nearby star-forming regions and all the gas in their foreground ($0.20\arcsec \times 0.46\arcsec$ for NIRSpec), whereas throughout the paper we compute column densities using narrow beams tracing the line of sight to star-forming regions.
To test the impact of this, we also compute column densities within cylindrical apertures of diameter $\sim 0.4\arcsec$ ($\sim2$~kpc at $z=8$), measuring the mean gas density across the entire aperture.
We find that the resulting column density distributions are not significantly affected by using a wider aperture, the main effect being that the typical standard deviations of the \NHI distribution are lower, going from an typical value of $\sigma_{\log \NHI} \sim 0.8$ to $\sim 0.4$.
We note this is a conservative estimate, as we are averaging equally all the gas within the aperture. In reality, the observed spectrum will be dominated by emission (and thus absorption) from the peak of the star-forming region, and the observed absorbing column density would be closer to a luminosity-weighted average over the aperture -- more similar to our narrow beam approach. Moreover, at $z>6$ typical UV sizes of galaxies have been shown to be smaller than a single MSA shutter \citep[$R_\mathrm{eff}\simlt400\,$pc, $\simlt0.08$\,arcsec, e.g.,][]{Yang2022,Morishita2024_size}, ensuring that all the emission from the star forming regions should be captured for sources well-centered in the MSA. Therefore, the impact of spatial apertures does not impact our conclusions.

\subsection{The nature of extremely strong galaxy-DLAs and predictions for observations} \label{subsec:disc_DLAs}

Recent JWST observations have uncovered a small fraction of $z\gtrsim5$ galaxies with extremely strong \Lya break turnovers. If interpreted as \Lya damping, albeit with uncertainties in UV continuum emission modelling, these may correspond to column densities $\NHI \simgt 10^{22}$\,cm$^{-2}$ \citep[e.g.,][]{Hsiao2024,Heintz2023b,Chen24,Mason25}. Such high column densities are comparable to those seen in the strongest GRB-DLAs at $z\sim2-6$ \citep[e.g.,][]{Jakobsson2006,Watson2006,Tanvir19,Saccardi2025} and so-called ``Extremely Strong'' DLAs (esDLAs) in QSO spectra \citep[e.g.,][which have been shown to have similar metallicities and kinematics as GRB-DLAs at $z\sim2$]{Noterdaeme2015,Ranjan20, Krogager2024}, both of which are expected to trace gas in the inner ISM \citep[$<1\,\rm kpc$,][]{Prochaska07,Pontzen10}. 
Our results suggest these high column densities inferred at $z>5$ also arise in the ISM ($\lesssim 500$\,pc, see Figure~\ref{fig:fractions_vs_r}). Thus, we consider it likely that the extremely strong galaxy-DLAs detected with JWST are also mostly associated with dense gas in the inner ISM. Furthermore, we find $\NHI \simgt 10^{22}$\,cm$^{-2}$ DLAs are predicted to be more common for more massive halos: $\sim30\%$ of sightlines to $M_h\gtrsim 10^{11}\,\rm M_\odot$ halos intersecting such high column density gas, compared to $<10\%$ for $M_h\lesssim 10^{10.5}\,\rm M_\odot$ halos.

If the JWST-detected strong galaxy-DLAs trace the inner ISM of massive halos, this should provide testable predictions for the environment of these sources and their spectra.
High-mass halos typically reside in the densest regions of the cosmic web, at the intersection of multiple filaments. These environments enable efficient gas and dark matter accretion, leading to frequent mergers, especially at high redshift. As a result, galaxies in more massive halos are expected to be surrounded by a larger number of satellites and companion galaxies. The number of nearby galaxies could therefore serve as a tracer of halo mass, potentially providing a way to test its correlation with $\NHI$.  

We analyze the number of neighboring galaxies within $\lesssim 100$~kpc of each system in the \code{SERRA} sample. We define neighbors as galaxies located at distances between $5-50$~kpc and brighter than $M_{\rm UV} \lesssim -19$, ensuring their detectability in JWST surveys \citep[e.g.,][]{Gelli21}.  
We find that stronger absorbers are more likely to have more neighboring galaxies: all galaxies with more than 4 neighbors have a median $\NHI > 10^{21.5} \,\mathrm{cm}^{-2}$.   

Interestingly, recent JWST results by \citet{Chen24} report the discovery of three strong DLA candidates at $z \sim 8$ in a protocluster of more than ten spectroscopically confirmed galaxies spanning $\lesssim 60$~kpc, estimated to have a halo mass $> 4\times10^{11}\,\rm M_\odot$ \citep{Morishita2022b}. By estimating \NHI from photometry, \citet{Witten2025} find that the galaxies in this protocluster with the highest inferred \NHI are in a small region of the field, $\lesssim 5 \, {\rm kpc}$, consistent with our picture that high \hi\ column densities arise in the inner regions of massive halos. Similarly, \citet{Mason25} find that five out of six robust DLA candidates at $z \sim 5.5 - 6$ in GOODS-S have spectroscopically confirmed close neighbors within $\lesssim 500$~kpc, suggesting that strong DLAs trace the most massive halos.
Currently, observational constraints are limited by the completeness of JWST spectroscopy. More complete spectroscopy around strong DLA candidates should be able to systematically test the connection between the prevalence of high column density systems and galaxy halo mass.

Another key prediction for the ISM origin of strong DLAs is the presence of metal absorption lines. In Sec.~\ref{subsec:one_galaxy}, we have shown the metal distribution is highly inhomogeneous in the ISM: gas is efficiently enriched by ongoing star formation and metals are driven outwards by SN outflows, while concurrent accretion of pristine neutral hydrogen from the CGM dilutes the gas inflowing through filaments.
However, in the immediate vicinity of the star-forming regions ($\lesssim 500$~pc), the metallicity always exceeds $Z\gtrsim 0.01 \rm Z_\odot$ and can even reach almost solar values. Beyond $\sim5\,{\rm kpc}$, metallicity drops to the simulation floor ($Z=10^{-3}\rm Z_\odot$) and the filamentary and clumpy CGM structures can be considered entirely pristine.

This stark contrast between the inner ISM and CGM, where the former is enriched and the latter is considerably lower metallicity, has direct implications for the nature of strong DLAs observed at high redshift. If the observed \highz DLAs predominantly trace the dense gas in the ISM star-forming regions, we expect the presence of a non-negligible amount of metals, traced by low ionization metal absorption lines at redshifts consistent with the systemic redshift of the galaxy. 
This would be consistent with GRB-DLAs and extremely strong QSO-DLAs, which have been shown to trace higher metallicity gas than typical QSO-DLAs \citep{Prochaska07, Ranjan20} and are thought to trace the inner ISM. Interestingly, especially high, near to solar, metallicities have been inferred for the highest column density GRB-DLAs with $\NHI > 10^{22}$\,cm$^{-2}$ \citep{Prochaska09, Kruhler2013}.

High-resolution spectra could directly test this for high-$z$ galaxies by detecting metal absorption lines. Low ionization interstellar absorption lines have been detected in correspondence with high density \hi\ in $z\sim 2 - 5$ galaxies \citep[e.g.][]{Reddy2016, Pahl2023}, indicating that metal enriched gas is more abundant for galaxies with larger \hi\ covering fractions, and our predictions shows a similar trend is expected beyond $z>6$.
We estimate line-of-sight metallicity for high column density sightlines with $\NHI > 10^{22}$\,cm$^{-2}$ in our sample, finding that the average mass-weighted gas metallicity is $Z \gtrsim 0.2~\rm Z_\odot$. Assuming solar abundance ratios, this corresponds to column densities of the key elements like C, Si, O, and Fe of $14 \lesssim \log N_X / \rm cm^{-2} \lesssim 17 $. Such high column densities should result in significant absorption from low ionization lines such as CII, SiII, OI, FeII, which are expected to be saturated and with equivalent widths of the order of $\rm EW= 1-5~\angstrom$ \citep[e.g.,][]{deUgarte-Postigo12}. The prospects for obtaining constraints on these absorption lines in individual  $z\simgt6$ galaxies are promising with JWST. $\rm EW \approx 1-5~\angstrom$ absorption lines have been detected in both stacked $R\gtrsim1000$ $z>6$ NIRSpec spectra \citep{Glazer25}, and individually in a handful of intrinsically bright and gravitationally lensed sources \citep{Boyett2023,Topping24,Topping2025a}.
Deep $R\gtrsim1000$ resolution spectroscopy of $z\simgt6$ sources with JWST should thus be able to confirm if strong galaxy-DLAs trace the inner ISM.

A potential caveat of our analysis concerns the modeling of H$_2$ formation and the consequent relative abundance of \hi\ and H$_2$ in the ISM, which introduces uncertainty in the abundance of the highest \NHI ($\simgt 10^{22}$\,cm$^{-2}$) sightlines. Due to self-shielding from UV photons, at such high column densities, a high fraction of \hi\ is predicted to be converted to molecular hydrogen \citep{Schaye2001b,Gurman2025}. The relative ratio of H$_2$ to \hi\ depends on the local UV radiation (which photo-dissociates H$_2$) and the dust mass and grain size (which catalyses H$_2$ formation). In \code{serra} the catalysis of H$_2$ on dust is independent of the grain size \citep{Jura74}; in principle a larger grain size distribution \citep[as recently suggested at high redshift, e.g.,][]{Markov2025, McKinney2025} may impact the efficiency of H$_2$ formation \citep[see e.g.,][]{Soliman2024}. We leave further exploration of this to future work.

\subsection{Redshift evolution in the ISM and CGM} \label{subsec:disc_CGM}

In Section~\ref{subsec:z_evolution}) we found an increase in the median CGM ($>2.5$\,kpc) column density, scaling approximately with cosmic density, $\NHI \sim (1+z)^2$, over $z\sim6-10$.
This is consistent with findings from other simulations which have found an increase in the covering fraction of LLS in the CGM with increasing redshift \citep[e.g.,][]{Rahmati15,Tortora24}. This likely reflects increasing cosmic gas densities and gas accretion rates, though as we noted above the covering fraction of low density gas in the CGM can be sensitive to feedback and resolution effects \citep{Rahmati15,Faucher-Giguere15,Bennett2020,Kimm2025}. 

By contrast, we find no strong redshift evolution in ISM column densities for a given halo mass where the highest column density gas ($\NHI \gtrsim 10^{20.5}$\,cm$^{-2}$), which produces damping wing absorption, arises.
Such weak evolution in ISM column densities may be expected if these are primarily driven by collapsed gas clouds, whose properties higher-resolution simulations have found to be nearly universal across cosmic time \citep{Guszejnov20,Wang2025}. 
The weak increase we find in the median column density ($\NHI \sim (1+z)^{0.3}$) is also close to what would be expected based on estimates of electron densities in galaxies' star forming regions with JWST \citep[$n_e\sim(1+z)^{1.5}$, e.g.,][]{Isobe2023,Topping2025}. 
The relationship between electron densities and HI densities is complicated due to the formation of HII regions. However, if the majority of HI absorption occurs in the ISM around massive stars, a better understanding of ISM densities and ionization, as well as selection effects which may increase the visibility of sources with high ISM densities \citep[thought to be linked to strong star formation bursts, boosting UV luminosity][]{Topping2025a}, will be important for interpreting DLAs in galaxy spectra.

Overall, as discussed by \citet{Mason25}, the stronger $\sim (1+z)^2$ redshift evolution in the CGM is still not sufficient to explain the observed increase in \Lya damping at $z\gtrsim6$, implying the increasingly neutral IGM plays a major role in shaping $z>6$ spectra.

We note our simulations only consider local radiation, and do not include a UV background. Gas at densities $n > 10^{-2} \rm cm^{-3}$ is expected to be self-shielded from the UVB \citep[e.g.,][]{Gnedin10, Rahmati13, Chardin2015}, and at high redshift, the local radiation field from local star-forming regions dominates over the background. 
Thus, we expect our results for high column density gas to be minimally affected by the lack of UVB.
Our results in the CGM are likely to be conservative -- as the inclusion of the UVB would decrease the HI fraction in lower density gas \citep[e.g.,][]{Schaye2001,Bolton2013,Nasir2021}.
Therefore, this approximation should not significantly affect our conclusions about the origin of galaxy-DLAs.

\section{Conclusions}\label{sec:conclusions}

In this work, we have used high-resolution cosmological zoom-in simulations from the \code{serra} suite to investigate neutral hydrogen in the ISM and CGM of galaxies during the Epoch of Reionization. Motivated by recent low resolution JWST spectroscopy showing damped \Lya absorption in $z>5$ galaxies, where local HI absorption can be hard to disentangle from absorption by the increasingly neutral IGM, our goal was to investigate the location and evolution of dense local (ISM+CGM) neutral gas in and around \highz galaxies with redshift and galaxy properties.
We analyzed $\sim 100$ galaxies at $z \sim 6 - 9.5$.
Our main findings can be summarized as follows:
\begin{itemize}
    \item The neutral gas distribution of \highz galaxies is complex and irregular in both the ISM -- continuously shaped by bursty star formation through feedback and frequent mergers and accretion -- and the CGM, characterized by dense pristine HI gas in filaments accreting onto the galaxies, and numerous HI dense clumps.
    \item By generating mock sightlines towards star-forming regions in galaxies, we find the distribution of HI column densities, \NHI, to individual galaxies is very broad ($0.5-1.5$~dex), due to strong sightline-to-sightline variations. We find typical median column densities, $\log \NHI/\rm cm^{-2} \sim 21-22$, comparable to the observed GRB-DLA \NHI distribution at $z\sim2-6$ \citep{Tanvir19}, with $\sim30\%$ of sightlines classified as strong DLAs $\log \NHI/\rm cm^{-2} > 22$, comparable to that inferred from JWST spectra \citep{Mason25}.
    \item We find dense neutral gas in the ISM, close to the emitting star-forming regions, $<1$\,kpc, is the dominant source of high column density absorption ($\log \NHI/\rm cm^{-2} \gtrsim 21$), with negligible contributions from the CGM or intervening proximate absorbers. For all galaxies across mass and redshift, the probability of intersecting a $\log \NHI/\rm cm^{-2} \gtrsim 21$ sightline within $<0.25$~kpc is $>50\%$, decreasing rapidly at larger distances.
    \item The median \NHI\ increases with halo mass, $\NHI \sim M_h^{0.38}$, as expected if \NHI\ scales with the virial radius ($\NHI \sim R_\mathrm{vir} \sim M_h^{1/3}$) reflecting a more extended CGM and the capability of deeper potential wells to accrete and retain more gas, adding to the dominant ISM HI column densities.
    \item We find no significant redshift evolution of \NHI distributions towards star-forming regions at fixed halo mass across the EoR, though we find a mild increase in CGM column densities scaling with mean cosmic density, as seen in other simulations.
    \item We investigate the origin of extremely strong galaxy-DLAs ($\log \NHI/\rm cm^{-2} \gtrsim 22$) detected with JWST.
    We predict these are associated with dense ISM gas along the line of sight to star-forming regions in massive halos ($M_h \gtrsim 10^{11} M_\odot$). We show the number of neighboring galaxies can provide a probe of this association with massive halos, as systems with at least 4 neighboring galaxies within 100\,kpc are predicted to typically have $\log \NHI/\rm cm^{-2} > 21.5$. Furthermore, we show that the absorbing gas in the ISM is expected to be metal enriched and produce low ionization metal absorption lines reaching EW~$\sim 1-5~\angstrom$~.
\end{itemize}

Our results show that \highz galaxy-DLAs in \code{SERRA} have HI column densities comparable to those of $z \sim 2 - 4$ GRB-DLAs. Our results imply that observed galaxy-DLAs likely trace neutral gas located close ($<0.25$~kpc) to star-forming regions. This motivates future higher-resolution simulations to uncover the impact of possible pc-scale structures in the HI distribution in the ISM, and higher resolution spectroscopy to associate strong galaxy-DLAs with their host galaxies via low ionization metal absorption lines.

Our results demonstrate that strong sightline-to-sightline variation is expected in the HI column density distribution for individual galaxies, and we predict little evolution in the distribution from $z \sim 6$ to $z \sim 10$. This suggests it may be challenging to identify individual sources affected by galaxy-DLAs based on galaxy properties alone when using only low resolution spectra ($R<1000$). However, it implies that \NHI\ distributions derived from spectra near the end of reionization at $z\sim5-6$, where the IGM has a minimal impact on \Lya damping wings, could provide valuable priors to interpret JWST spectra probing the earliest stages of reionization at $z>6$, similar to approaches taken for interpreting the evolution of \Lya emission during reionization \citep[e.g.,][]{Treu2012,Mason2018,Tang2024,Tang2024c,Nakane24}.

\begin{acknowledgements}
We thank Taysun Kimm for valuable discussions and comments on a draft of the manuscript, Tucker Jones and Naveen Reddy for providing comments on a draft, and Dan Stark for useful discussions.
VG acknowledges support by the Carlsberg Foundation (CF22-1322).
CAM acknowledges support by the European Union ERC grant RISES (101163035), Carlsberg Foundation (CF22-1322), and VILLUM FONDEN (37459). Views and opinions expressed are those of the author(s) only and do not necessarily reflect those of the European Union or the European Research Council. Neither the European Union nor the granting authority can be held responsible for them.
The Cosmic Dawn Center (DAWN) is funded by the Danish National Research Foundation under grant DNRF140.
This work has received funding from the Swiss State Secretariat for Education, Research and Innovation (SERI) under contract number MB22.00072.
We acknowledge the CINECA award under the ISCRA initiative, for the availability of high performance computing resources and support from the Class B project SERRA HP10BPUZ8F (PI: Pallottini).
We gratefully acknowledge computational resources of the Center for High Performance Computing (CHPC) at SNS.
\end{acknowledgements}

\bibliography{refer,codes,library_cm}
\bibliographystyle{aasjournal}

\end{document}